\documentclass[preprint]{elsarticle}
\usepackage{lineno,hyperref}
\usepackage[utf8]{inputenc}
\usepackage{graphicx}
\usepackage{graphics}
\usepackage{amsmath}
\usepackage{float}
\usepackage{siunitx}
\usepackage{natbib}
\usepackage{blindtext}
\usepackage[nice]{nicefrac}
\usepackage{lscape}
\usepackage{csquotes}
\usepackage{enumitem}
\newcommand*{\ditto}{\texttt{"}}

\newcommand{\expectedk}[1]{\ensuremath{\left<k\right>\!\left(#1\right)}}

\graphicspath{{./figures/}}
\journal{arXiv}
\begin{document}
\begin{frontmatter}
\title{Face-masks save us from SARS-CoV-2 transmission}

\author[1]{Gholamhossein Bagheri\corref{mycorrespondingauthor}}
\ead{gbagher@gwdg.de}
\author[1]{Birte Thiede}
\author[1]{Bardia Hejazi}

\author[1]{Oliver Schlenczek}

\author[1,2,3]{Eberhard Bodenschatz\corref{mycorrespondingauthor}}
\cortext[mycorrespondingauthor]{Corresponding author}
\ead{eberhard.bodenschatz@ds.mpg.de}

\address[1]{Max Planck Institute for Dynamics and Self-Organization (MPIDS), Göttingen 37077 , Germany}
\address[2]{Institute for Dynamics of Complex Systems, University of Göttingen, Göttingen 37077 , Germany}
\address[3]{Laboratory of Atomic and Solid State Physics and Sibley School of Mechanical and Aerospace Engineering, Cornell University, Ithaca, NY 14853, USA}

\begin{abstract}
We present results on the infection risk from SARS-CoV-2 under different scenarios based on measured particle size-dependent mask penetration, measured total inward leakage, measured human aerosol emission for sizes from 10nm to 1mm, and re-hydration on inhalation. Well-mixed room models significantly underestimate the risk of infection for short and direct exposure. To this end, we estimate the upper bound for infection risk with the susceptible in the infectious exhalation cloud or wearing masks by having the masked susceptible inhale the entire exhalation of a masked infectious. Social distances without a mask, even at 3m between speaking individuals results in an upper bound of 90\% for risk of infection after a few minutes. If both wear a surgical mask, the risk of infection for the person speaking remains below 26\% even after 60 minutes.  When both the infectious and susceptible wear a well-fitting FFP2 mask, the upper bound for risk is reduced by a factor of 60 compared to surgical masks. In both cases, face leakage is very important. For FFP2 masks, leakage is low in the nasal region and directed upward, which can be further reduced significantly by applying double-sided medical tape there. Considering that the calculated upper bound greatly overestimates the risk of infection, and the fact that with a poorly worn mask even the upper bound we calculated is very low, we conclude that wearing a mask, even with some leakage, provides excellent third party and self-protection. 
\end{abstract}

\end{frontmatter}

\section{Introduction}
Infectious airborne diseases such as the Severe Acute Respiratory Syndrome (SARS) 2002, the Avian Influenza and Swine Influenza (H1N1), and more recently, the Coronavirus disease (COVID-19) are transmitted via direct and indirect exposure from an infectious to a susceptible \cite{Zhang2020,WHO2020, Nordsiek2021, Pohlker2021}.
One route of transmission 
is via airborne transport of respiratory-origin particles released from the respiratory tract of an infectious. 
(In this study, we use the terms \emph{particle(s)} to refer to \textless\SI{100}{\micro\meter} particulate matter suspended in air, regardless of composition.)
Human exhaled particles (also referred to as aerosols or droplets) greatly vary in composition and size, and span several decades in length scale \cite[e.g. see][and references therein]{Nordsiek2021, Pohlker2021}.
Concentrations of exhaled particles and their size have been found to strongly depend on the type of respiratory activities, for example speaking, or singing as compared to breathing.
Vocalisation-related respiratory activities, i.e. sound pressure, peak airflow frequency and articulated consonants, strongly influence particle emission as summarized in P\"ohlker et al. \cite{Pohlker2021}.
Particles may contain single or multiple copies of pathogens when exhaled by the infectious, and when inhaled by the susceptible, there is a risk of infection given the absorbed infectious dose \cite{Nordsiek2021}.
Furthermore, relative humidity and temperature influences the drying and settling of particles by gravity \cite[see][and references therein]{Nordsiek2021, Pohlker2021}.
There is an ongoing debate about whether COVID-19 is transmitted primarily via aerosols or droplets \cite{GREENHALGH20211603, Leung2021}. There is also a longstanding debate about what is meant by \emph{aerosols} and \emph{droplets} \cite{randall2021did}.
At their core, these debates are fueled by our inadequate understanding of how airborne disease transmission works, or simply put, how particles produced in the respiratory tract of the infectious become airborne and enter the respiratory tract of the susceptible. As simple as it sounds, the detailed  mechanisms involved in each part of this process are extremely complicated \cite{Pohlker2021}. 

On the source side, i.e. the infectious,  we have the physio-anatomical dependency  intertwined with the complexity of particle production controlled by the respiratory maneuver and the size-dependent pathogen concentrations due to differences in their origin sites and/or their volume and/or the nature of the pathogen itself.
In the air we have the turbulent \emph{cloud} exhaled by the infectious source while being turbulently diffused and advected into the ambient.
The advection and turbulent diffusion of the exhale cloud is strongly influenced by the ambient thermodynamical conditions and the airflow, e.g. the type and strength of the ventilation in indoor environments, or outdoors, as well as in large indoor spaces the air currents that carry away the exhale \cite{Hejazi2021}.
While being advected, particles may lose their volatile content due to evaporation and shrinkage, which is determined by their chemical composition, ambient thermodynamical conditions, exhale flow velocity, mixing with the ambient air, and the time they need to reach an equilibrium size \cite{Chong_2021,Pohlker2021}.
While being advected with the flow some particles will be lost due to deposition on nearby surfaces depending on their size, shape, and density and might get re-suspended at a later time.
In addition, pathogens might loose their infectiousness before being absorbed by the susceptible receptor.
On the receptor side, i.e. the susceptible, the mechanisms controlling the inhalability and absorption of pathogen-laden particles depend not only on the physical-anatomical properties of the receptor, but also on the receptor's breathing maneuver, inhaled particle size and composition, their re-hydration growth rate due to condensation in the receptor's airways, and the temperature and relative humidity of the inhaled air.
This is a multi-scale problem and physico-chemical processes reminiscent of the challenges in atmospheric clouds, with striking similarities - and differences. 

Atmospheric clouds and human exhalation clouds contain particles with similar size distributions. The water droplets and aerosols in atmospheric clouds are transported by turbulent flow and are mixed and diluted with atmospheric air.
Depending on the entrained air within the cloud, the droplets can also decrease or increase in size due to evaporation/condensation \cite{Bodenschatz2010}.
Furthermore, similar to cloud droplets, human saliva and epithelial mucosal fluid contain various salts that act as electrolytes and buffers, e.g., NaCl, and thus can be highly hygroscopic. Although there are many important differences between exhalation and atmospheric clouds, their similarities and the difficulty in understanding the microphysics of clouds lead to an unfortunate conclusion, namely: it will be extremely difficult  to fully understand the physics or predict its effects with an acceptable degree of certainty.
Uncertainties about airborne disease transmission were the main reason for the differences in hygiene measures implemented in different countries. A notable example is the use of face masks, which were initially discouraged to the public or recommended only for symptomatic cases and health care workers \cite{Greenhalgh2020}. Another example is the debate about the effectiveness of social distancing in reducing the risk of transmission \cite[see, e.g.,][]{Chu2020, Bazant2021}. Understanding the efficacy of face masks is essential to better understand the source-medium-receptor problem. It is challenging because it introduces additional nontrivial parameters into the calculation of infection probability, namely mask filter penetration and face seal leakage around the mask \cite{hinds1987performance, shaffer2009respiratory, holton1987particle, rengasamy2014quantitative} and also the change in exhalation flow through the mask \cite{Verma2020, Ishii2021}.

During the COVID-19 pandemic, good progress was made on the \emph{medium} problem in the source-medium-receptor trilogy by greatly simplifying the dispersion of infectious aerosols using the \emph{well-mixed room} assumption\cite[e.g.][]{Lelieveld2020, BUONANNO_2020, Miller_2020, Nordsiek2021, Bazant2021,Jimenez_COVID19app_CIRES_2020,heads}  In well-mixed room models, the pollutants (i.e., exhaled particles) are assumed to be diluted and completely mixed throughout the room volume before they reach the receptor. As a result, the concentration of particles in the room is the same (instantaneously) at all locations in the room and decreases exponentially with time, which depends on the room air exchange rate, deposition rate, filtration rate, and particle sizes. With this assumption, it is possible to study the mean properties in a room without having to resolve to turbulent fluctuations or location dependencies.

Particle deposition and mixing with uncontaminated air in the room can be modeled with well-established analytical and empirical models \cite[e.g. see][and references therein]{riley2002indoor, Nazaroff_1998, Nicas_2005, Lai_2000, He_2005, Nazaroff2008}.  There has been considerable success in assessing the risk of infection in well-mixed rooms to understand, albeit crudely, the relative risk of infection in different social settings and spatial conditions \cite{BUONANNO_2020, Miller_2020, Nordsiek2021, Bazant2021}. In practice, however, there will be concentration fluctuations in a room even if the air flow in the room is a fully developed turbulent flow. Since the exposure time scales of the receptor are usually much larger than the characteristic time scales of the turbulent concentration fluctuations, these can be approximated quite well by average concentrations. So what are the main problems with the well-mixed assumption?  It is that it breaks down when the distance between source and receptor decreases and/or when the room volume increases. This question has been addressed in numerous studies, e.g., by dividing a space into near-field and far-field regions \cite{Cherrie_2011, Nicas_2008, Arnold_2017} or by detailed numerical simulations of specific scenarios \cite{Sajjadi2016,Gao2008,Zhao2009}.

Here, we report an analysis of airborne disease transmission, specifically COVID-19, based on new data and models combined with improvements to established models and include 
\begin{enumerate}[topsep=1ex,itemsep=-1ex,partopsep=1ex,parsep=1ex]
\item{ the human exhalation particle database published in \cite{heads},} 
\item{ the total inward mask penetration measured on human subjects in this study,} 
\item{ the localized aerosol dispersion in large indoor spaces from recent investigations of \cite{Hejazi2021},}
\item{ the mono-/poly-pathogen infection risk model \cite{Nordsiek2021},} 
\item{ the established respiratory tract deposition model \cite{ICRP1994} refined to include time-dependent hygroscopic growth for particles entering the respiratory tract,}
\item {the particle shrinkage due to evaporation.} 
\end{enumerate}

\noindent With this we answer two questions:

\begin{itemize}
\item{ What are the environments in which well-mixed room assumptions fail and near-filed exposure should be taken into account for estimating the infection risk?}
\item{ What is the influence of face-masks and social distancing on SARS-CoV-2 airborne transmission?}
\end{itemize}

\noindent While similar questions were asked in various forms individually or together, and in different contexts in previous important studies, such as exposure to secondhand smoke, hazardous substances, or airborne pathogens, e.g., \cite{Klepeis1999, Nazaroff2008, Licina2017, Lelieveld2020, Yang2020,Abkarian2020, BUONANNO_2020, Pohlker2021, cheng2021face} our work builds upon and extends these earlier works. 

\section{Results}
\subsection{Mask Efficacy}
 Total Inward Leakage ($TIL$) is defined as the sum of penetrations during inhalation through the mask fabric, $P_{in}=P_{filter}$ and the \emph{face seal leakage} $L_{in}$, i.e. $TIL = P_{in}+L_{in}$. We have measured filter penetration $P_{filter}$ and total inward leakage $TIL$ in this study. The results for these measurements are presented below.\\
\subsubsection{Filter penetration}
\begin{figure}[htbp]
	\centering
	\includegraphics[width=1\linewidth]{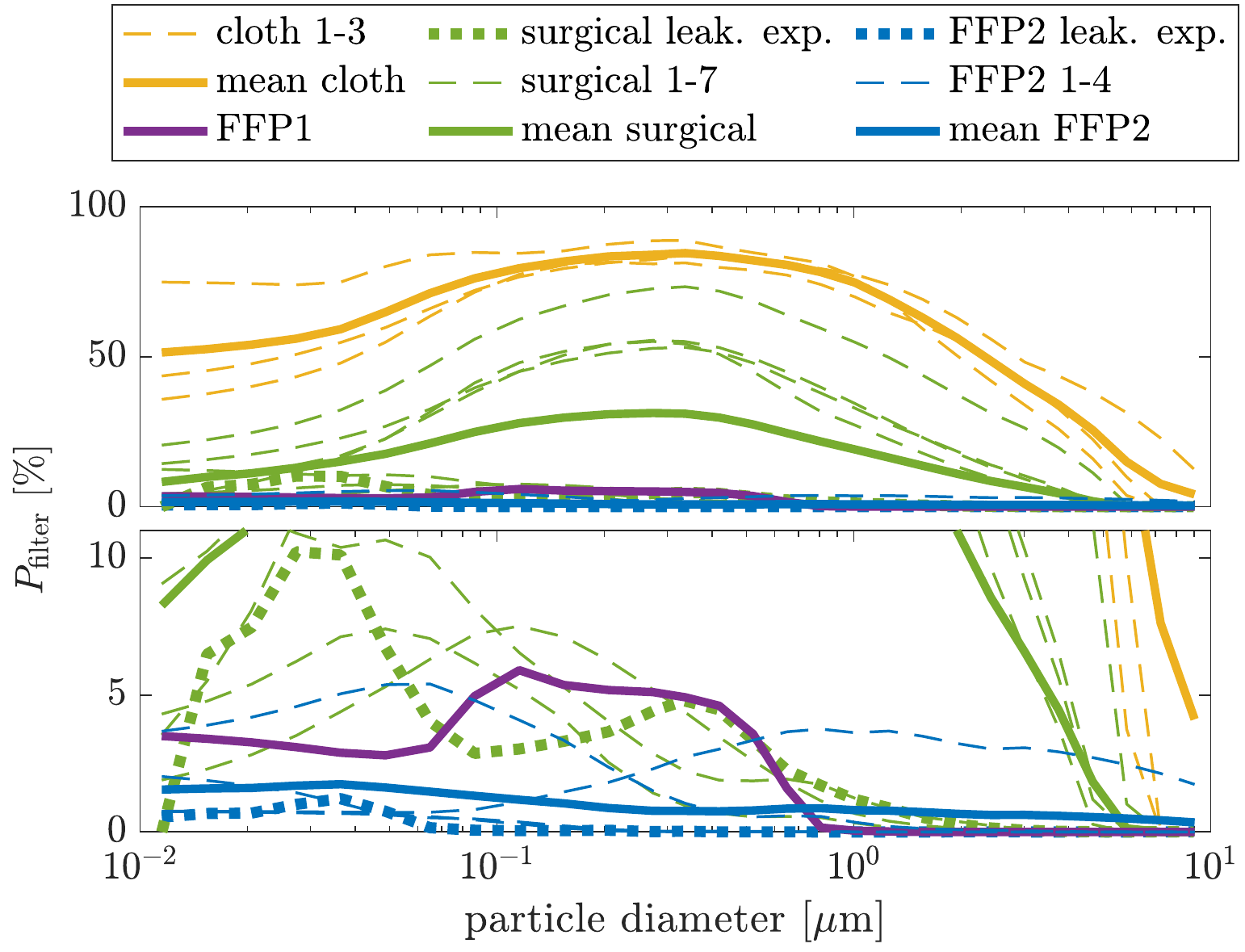}
	\caption{Particle penetration as a function of particle size through the filter material of the tested cloth, surgical, FFP1, and FFP2 masks. Dashed lines show the results from measurements of filter penetration of different masks in setup 1. Dotted lines show the results from filter penetration measurements in setup 2 for the two masks that were used in the leakage experiment. The solid lines are the mean of each mask type.}
	\label{fig:masktransmission}
\end{figure}

Figure \ref{fig:masktransmission} shows the measured filter penetration of three different cloth, eight surgical, one FFP1 and five FFP2 masks.
Between all the masks examined in this study, the cloth masks have the highest filter penetration with the maximum penetration for particles around \SI{0.3}{\micro\meter} diameter, whereas the FFP2 masks have the lowest penetration as shown in Fig. \ref{fig:masktransmission}. On average the cloth mask filters have a penetration of 85\% for particles with a diameter of \SI{300}{\nano \meter} with the worst-performing mask filter material reaching a penetration close to 90\%. These results are within the range of penetration values measured previously\cite{zangmeister2020filtration, morais2021filtration, konda2020aerosol,Drewnick2021}. All the examined FFP2 masks are below the 6\% limit set by the EN 149:2001+A1:2009 standard \cite{DinNorm}. On average the filter penetration further decreases with particle size for particles larger than \SI{50}{\nano \meter}, this tendency is much stronger for three of the tested FFP2 masks.

Surgical masks fall into two categories, four exhibit \textless 12\% filter penetration with maximum penetration for very small particles ($30-\SI{120}{nm}$) while the other four have filter penetration values almost as high as cloth masks. The latter category also shows the same \emph{characteristic maximum filter penetration} at \SI{0.3}{\micro\meter} as the cloth masks, ranging from 50\% to 75\%. The strongly varying results from previous studies \cite{zangmeister2020filtration, balazy2006n95, grinshpun2009performance, oberg2008surgical} agree with our finding that surgical masks exhibit vastly different filtering characteristics depending on the model.  

The dotted lines in Fig. \ref{fig:masktransmission} show the filter penetration of particles through the filter material of the FFP2 and surgical masks that were used in the leakage experiments measured with penetration setup 2.
The filter of the surgical mask has a maximum filter penetration of 10\% for particles of around \SI{30}{\nano \meter} diameter. A smaller peak in the penetration of about 5\% can be observed for particles with \SI{0.3}{\micro \meter} diameter. For larger particles the filter penetration further decreases.
The filter penetration of the FFP2 mask was smaller than 1.3\% for all particle sizes and below 0.03\% for particles larger than $\SI{30}{\nano \meter}$. For increasing particle size the filter penetration decreases drastically for both types of masks.
A more detailed discussion of the filter penetration results in comparison to existing literature can be found in the supplementary information (SI, section 2A).
It has to be noted that the flow velocities through the masks (and the equivalent breathing rates) vary by a factor of 4 between the two setups used in our penetration measurements. The flow rates in setup 1 are smaller than the one required by the European standard EN 149:2001+A1:200 (\SI{95}{\liter \per \minute}) \cite{DinNorm}. Despite filter penetration being reported to depend on breathing flow rates and therefore flow velocity through the material \cite[e.g.][]{he2013effect, he2014does, balazy2006n95, balazy2006manikin} our results do not seem to be noticeably influenced by it (dotted lines overlap with dashed lines, Fig \ref{fig:masktransmission}). The differences between different masks of a certain type exceed the potential influence of flow rate.

\subsubsection{Total inward (and outward) leakage}
\label{sec:TIL_results}
\begin{figure}[htbp]
	\centering
	\includegraphics[width=1\linewidth]{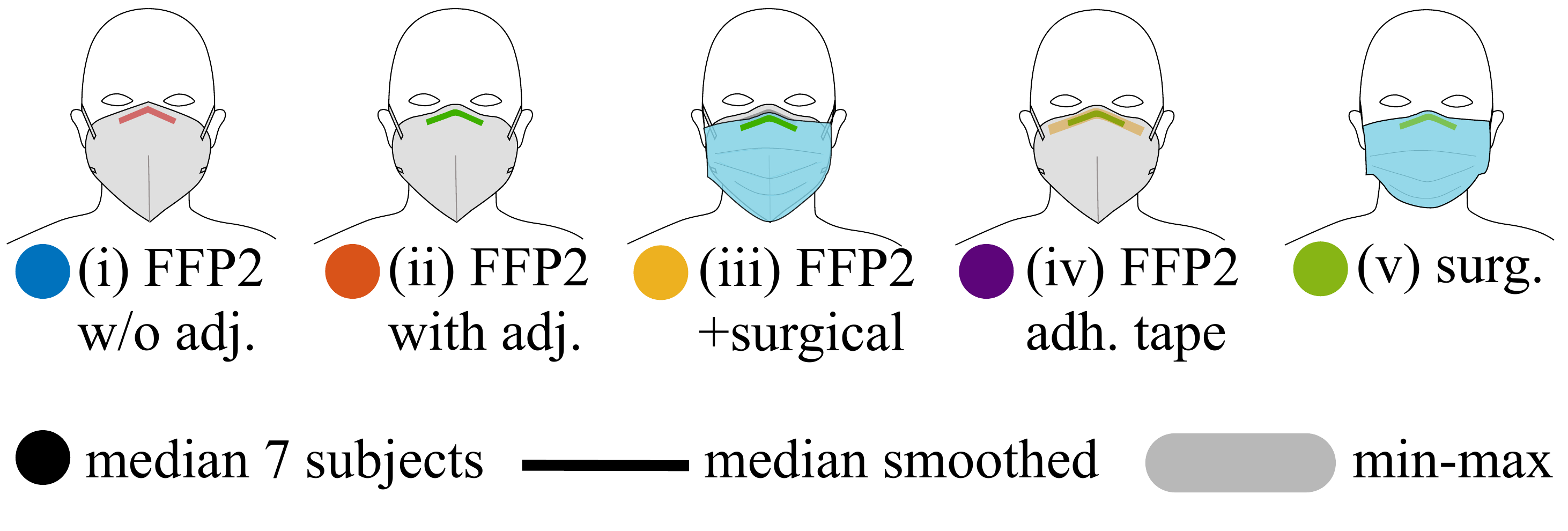}\\
	\includegraphics[width=1\linewidth]{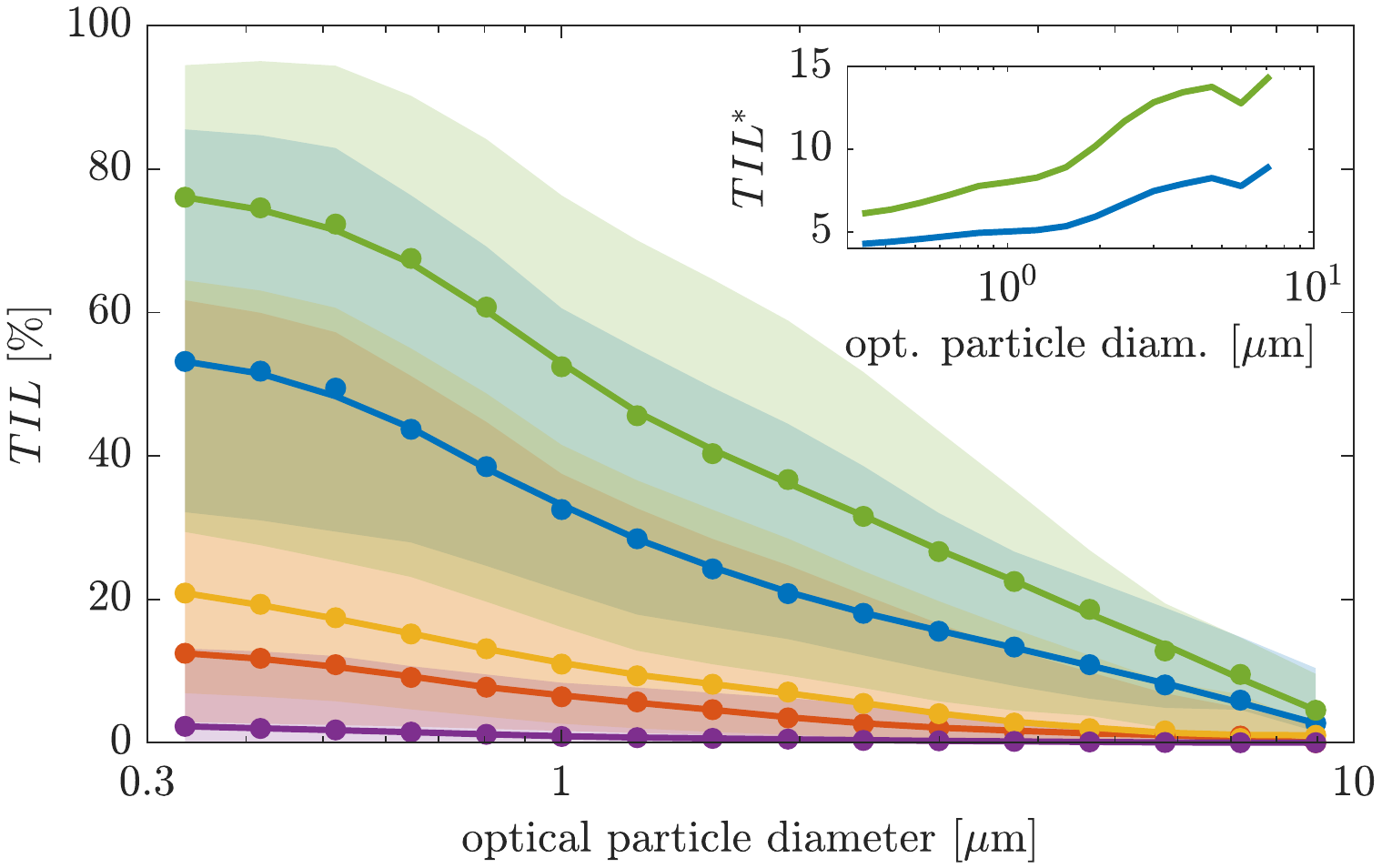}
	\caption{Median of the total inward leakage over all subjects for different mask-wearing cases. Smoothed curves are the 3-point moving average. Shaded areas show minimum and maximum as an indication of variability in total inward leakage for different subjects. The first-last bin total leakage values are: (i)53.2-2.7\% (ii)12.5-0\% (iii)20.9-1.0\% (iv)2.3-0\% (v)76.0-4.5\%. Inset shows the total inward leakage of the surgical mask and the FFP2 mask without adjustment normalized with the total inward leakage of the adjusted FFP2 mask $TIL^*=TIL/TIL_{FFP2,adj.}$.}
	\label{fig:maskleakage}
\end{figure}

 The median total inward leakage of seven subjects for mask cases (i) to (v) is shown in Fig. \ref{fig:maskleakage}. 
The shaded regions represent the range of leakage values from the worst-performing to the best-performing mask/subject combination. The total inward leakage decreases for all mask wearing cases (i) to (v) with particle size for particles larger than \SI{300}{\nano \meter}, which agrees with the literature \cite{hinds1987performance, willeke1992filtration, weber1993aerosol, cho2010large, lee2005respiratory, lee2008respiratory, grinshpun2009performance}. The best mask fit, i.e. least leakage, is found in case (iv), in which the leakage at the nosepiece of the FFP2 mask is eliminated by using the adhesive tape as explained in the methods section \ref{sec:til_methods}. This indicates the strong influence of nosepiece leakage, which agrees with infrared observations on N95 masks \cite{lei2013simulation} and results from simulation studies  \cite{lei2013simulation,peric2020analytical} and observations of tracer particles with half-mask respirators by Oestenstad \textit{et al.} \cite{oestenstad1990distribution}. 
In another investigation by Oestenstad \textit{et al.} \cite{oestenstad2010factors}, however, leaks at the cheeks were found to play a significant role as well, which can possibly be explained by considering the fact that facial dimensions play an important role on leak positions \cite{lei2013simulation, oestenstad2010factors}. Overall, our findings agree with the literature on total inward leakage but we find higher leakages for an adjusted mask than most other studies investigating mask leakage on human subjects. More detailed comparisons with existing leakage measurements on human subjects can be found in the supplementary information (SI, Fig. S11 and S12).

Due to the high filter efficiency of the FFP2 mask (i.e. as shown in Fig. \ref{fig:masktransmission}), it can be assumed that the majority of the particle concentration penetrating into the mask is caused by the face-seal leakage. Wearing an FFP2 mask without any adjustment, case (i), leads to total inward leakage of 53\% for the smallest particle bin (0.3 to $\SI{0.37}{\micro \meter}$), which decreases to 16\% for \SI{3}{\micro \meter}.   
By simply adjusting the mask nosepiece to the nose, case (ii), the mask's $TIL$ is improved by a factor of 4.3 and for the smallest particles by a factor of 7.5 for \SI{3}{\micro \meter} particles (see inset Fig. \ref{fig:maskleakage}).

The surgical mask, however, is associated with the highest total inward leakage with the maximum value being in excess of 70\% occurring for the smallest particle size. This is caused both through relatively high filter penetration (~5\% for particles around \SI{0.3}{\micro \meter} as shown in Fig. \ref{fig:masktransmission}) and the evidently high leakage. The total inward leakage of the surgical mask shows a similar trend to that of the FFP2 mask without adjustment and is 6 times higher compared to the adjusted FFP2 mask for the smallest particles and over 12 times higher for particles \textgreater \SI{3}{\micro \meter}. This suggests that an overall better fitting mask has an even higher relative protection from large particles compared to small particles.

Wearing an additional surgical mask on top of the adjusted FFP2 masks, case (iii), seems to have an overall negative effect on the total inward leakage compared to case (ii). The effect of double filtering is negligible as the filtration efficacy of the FFP2 mask is already large (\textgreater 99.98\% for particles \textgreater \SI{0.3}{\micro\meter}). A possible explanation for the decreased protection (increased inward leakage) could be that the additional pressure on the FFP2 mask caused by the surgical mask distorts the FFP2 mask and therefore causes new face seal leaks. For some individual subjects, however, the surgical mask on top of the FFP2 mask lead to a slight improvement, i.e. decreased total inward leakage. In our early experiments, in which no diffusion dryer was used, we have observed an overall improvement for case (iii) over case (ii). The results for wearing a surgical mask on top of an FFP2 mask are therefore inconclusive. 
Facial hair is found to increase the face seal leakage and therefore total inward leakage (SI, Fig. S10).

We found an increased $TIL$ when subjects were mouth-breathing compared to nose-breathing for cases (i) and (ii). With the mask taped to the nosepiece, case (iv), we observed the exact opposite (cases (iii) and (v) are not investigated for mouth vs nose breathing experiments, see SI, Fig. S8). Thus the impact of nose vs. mouth breathing on the leakage remains inconclusive. Reading with a loud voice ($\sim$ 80-90 dBA) is found to decrease the $TIL$ as compared to breathing through the nose by as much as a factor of 3 for the adjusted FFP2 mask (see SI, Fig. S5). As a result, leakage values measured during breathing experiments are most likely an upper estimate for activities that are not associated with significant movement of the mask relative to the face.
The total inward leakage is expected to decrease with higher inhalation rates and to not strongly depend on breathing frequency \cite{he2013effect}.

For particles with diameter 0.1-\SI{0.3}{\micro \meter}, smaller than what we measured, the total inward leakage can be expected to plateau \cite[cf.\ e.g.][]{lee2008respiratory, grinshpun2009performance}. For larger particles than measured, \textgreater \SI{10}{\micro \meter}, the total inward leakage can be assumed to drop to zero. Zero leakage median can already be observed for the largest measured particles with average diameter of \SI{9}{\micro \meter} for the adjusted and taped FFP2 mask (cases (ii) and (iv)).
Similar assumptions were used by Hinds \textit{et al.}\ \cite{hinds1987performance2}.
As explained in Materials and Methods, The Total Outward Leakage during exhalation $TOL=P_{ex}+L_{ex}$ can be reasonably assumed to be equal to $TIL$. 

\subsection{Effective respiratory tract penetration}
Particles undergo a complex journey after exhalation by the infectious until deposition in the susceptible respiratory tract.
The overall effects of outward and inward mask penetration, as well as the transient particle size that depends on environmental conditions and time since exhalation, must be considered in order to capture the true particle penetration from infectious to susceptible. 
At the time of exhalation, the particles are in an air parcel with a very high relative humidity of about 99.5\%. 
The ambient relative humidity is much lower and typically 40-70\%. 
As a result the particles shrink in size compared to the \emph{wet} diameter on exhalation. 
The shrinkage factor $w = d_0/d_e$, defined as the exhalation particle diameter by the infectious $d_0$ to the equilibrium diameter before inhalation by the susceptible $d_e$, influences the combined penetration from infectious to susceptible. 
For typical room conditions we expect $w\sim4$ as explained in the Materials and Methods. 
In the presence of particle shrinkage, the penetration curves through susceptible mask-fabric shown in Fig. \ref{fig:masktransmission} in inhalation $P_{in}$ is shifted to the right compared to that of the infectious mask-fabric in exhalation $P_{ex}$. 
The same is true for faceseal leakage during exhalation $L_{ex}$ and inhalation $L_{in}$.
As a result, the combined penetration in mask scenarios is rather complicated to estimate.

\begin{figure}[htbp]
	\centering
	\includegraphics[width=1\linewidth]{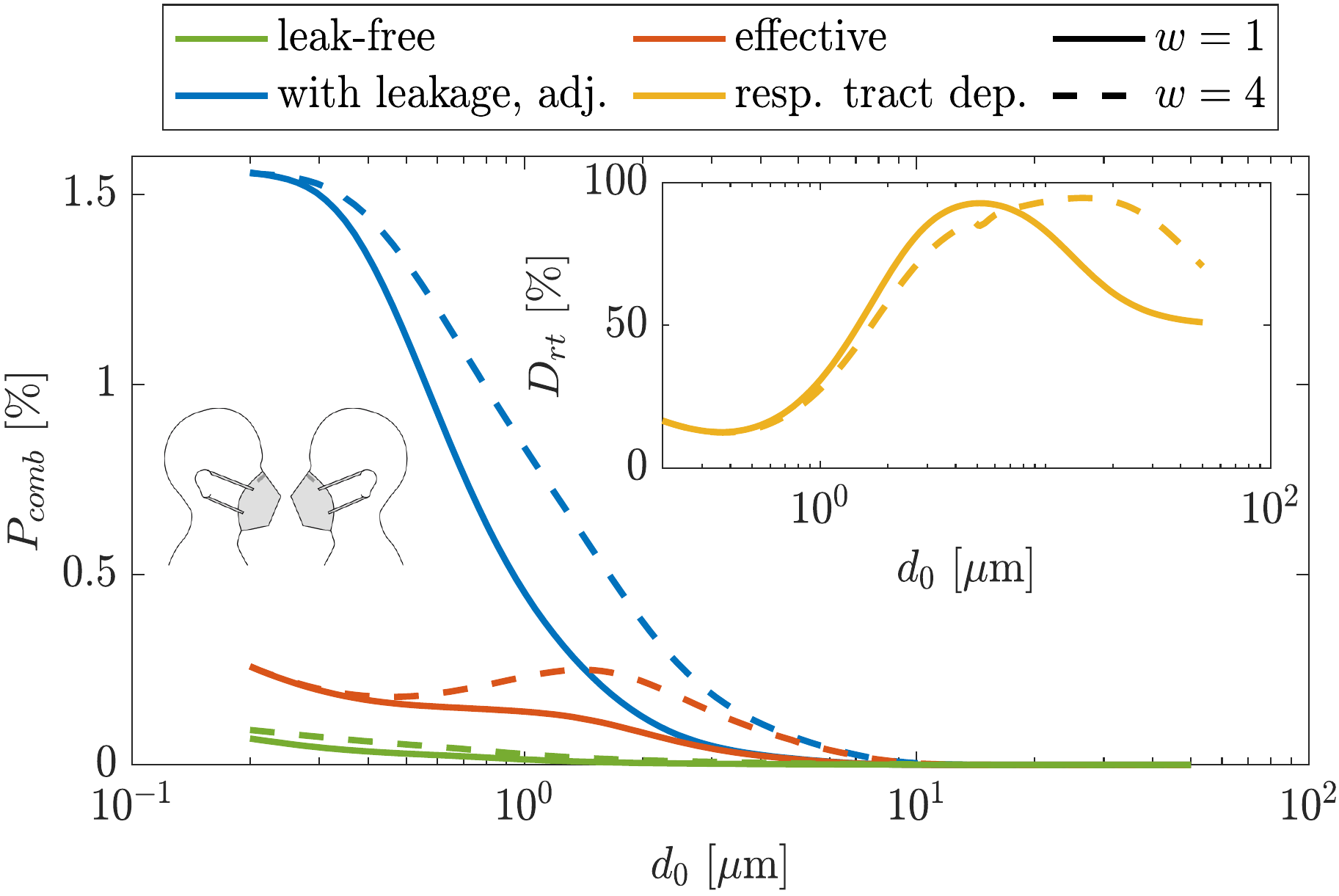}
	\caption{Combined penetration values when both infectious and susceptible are wearing FFP2 masks, i.e. mask-FF scenario (combined penetration for mask-SS scenario is shown in SI, Fig. S13), and at different shrinkage factors of $w = 1$, i.e. no shrinkage, and $w = 4$ as a function of particle diameter at exhalation, i.e. wet diameter. \enquote{Leak-free} curves correspond to $P_{ex} P_{in}$, \enquote{With leakage, adj.}
	curves correspond to $\left(P_{ex}+L_{ex}\right)\,\left(P_{in}+L_{in}\right)$ and \enquote{effective} curves correspond to $D_{rt}\left(P_{ex}+L_{ex}\right)\,\left(P_{in}+L_{in}\right)$. Respiratory tract deposition $D_{rt}$ is shown in the inset for $w=1$ and $w=4$.}
	\label{fig:effective_penetration}
\end{figure}

If we only consider ideal leak-free masks, the combined penetration through the fabrics of infectious and susceptible FFP2 masks $P_{in} P_{ex}$ is very low and is slightly affected by the shrinkage ratio as shown in Fig. \ref{fig:effective_penetration}. 
In practice, however, masks are always associated with leakage. 
The combined penetration of adjusted FFP2 masks with leakage taken into account, i.e. $\left(P_{in}+L_{in}\right)\,\left(P_{ex}+L_{ex}\right) = TIL\times TOL$, is much higher than that of ideal masks for the whole range of particle sizes as shown by the blue curves in Fig. \ref{fig:effective_penetration}.
The impact of shrinkage factor is also more visible for the leak-included than leak-free penetration.
The combined penetration for FFP2 masks with leakage increases with particle shrinkage factor since large particles that penetrate through the mask of infectious have higher penetration probability through the mask of susceptible as they get smaller.

However, for investigating the infection risk, the combined penetration should also take into account the deposition in the respiratory tract of the susceptible $D_{rt}$, which is shown in the inset of Fig. \ref{fig:effective_penetration}.
The $D_{rt}$ for the case when $w=1$ is based on the original model provided by the ICRP model \cite{ICRP1994}. 
For $w=4$, however, we have assumed particles undergo an unsteady hygroscopic growth and, hence, the deposition fraction in different regions of the respiratory tract behaves differently as shown by the dashed line in the inset of Fig. \ref{fig:effective_penetration} (for more details see Materials and Methods). 
Besides hygroscopic growth, particle shrinkage increase the inhalability probability of particles with initial wet diameter of \textgreater \SI{7}{\micro\meter}.
In addition, particles with wet diameters of \SIrange[]{1}{3}{\micro\meter} would have about 10\% lower probability of deposition if they shrink by a factor of 4 before inhalation. 

The combined effect of masks with leakage and the respiratory deposition, i.e.
effective penetration $=D_{rt}\,\left(P_{in}+L_{in}\right)\,\left(P_{ex}+L_{ex}\right)$, are also shown in Fig. \ref{fig:effective_penetration}. 
It can be seen that when a shrinkage factor of 4 is considered the maximum penetration occurs for \SI{1.5}{\micro\meter} particles, whereas for shrinkage factor of 1 (i.e. no shrinkage) the maximum penetration occurs for the smallest particle size. 
For surgical masks the results follows the same trends as shown in Fig. \ref{fig:effective_penetration} but the penetration magnitudes are much higher (see Fig S13 in SI).

\subsection{Well-mixed room vs. near-field exposure}

In order to determine the the most important factors that contribute to disease transmission via respiratory particles, we must identify the key elements involved in particle dynamics in a specific environment. 
As mentioned earlier, in a well-mixed environment concentration of particles decay exponentially in time as $exp(-t/\tau)$, where $\tau$ is decay time and $1/\tau$ is called the decay rate. 
The decay rate is also equal to the nominal Air Change per Hour (ACH), which is defined as the ratio of the room clean-air supply rate $Q$ to the room volume $V$ \cite{Nicas_1996}. 
Many experimental studies have investigated contaminant transport in indoor environments of various sizes and ventilation rates.
Table~\ref{table:lit} shows the details of some notable previous experiments studying particle dynamics in indoor environments and those measured in this study.
We extracted the decay times from contaminant concentration in time figures in studies that do not themselves directly report these values.

The available data in literature and the additional data from our experiments show that generally the contaminant decay rate is inversely related to the volume in which the contaminant is released and the ACH of the indoor space.
As the room size and ACH increase, the decay rate of contaminants decrease which can be seen in the study by Mage and Ott~\cite{Mage1996} in the large volume of a tavern and the study by Hejazi \textit{et. al.}~\cite{Hejazi2021} of aerosol transport in large hardware stores.
The decay rates measured in the large indoor space of hardware stores are size independent and $\tau$ is approximately constant for all measured particle sizes (0.3-\SI{10}{\micro\meter})\cite{Hejazi2021}. 
The small measured decay rates in large indoor spaces suggest that existing well-mixed contaminants in the environment do not play a critical role in infection risk calculations and that the strongest probability of infection is created by direct contact with an infected individual and their exhale particle emission.
In contrast, small, poorly ventilated rooms have extremely long decay times and particles remain airborne for long times. An example of long decay time in small spaces can be seen in the restroom measurements of Hejazi \textit{et. al.}~\cite{Hejazi2021} where $\tau$ is dependent on particle size.
In such small spaces, fine particles linger in the environment for longer times and even large particles ($>$\SI{5}{\micro\meter}) have considerably long residence times as compared to well ventilated large spaces. 

Some of the experiments in Table~\ref{table:lit} do not follow the general observation for the relationship between $\tau$ and room size and ventilation rate. These experiments are special scenarios that differ from the other reported measurements. The experiments by Ott \textit{et. al.}~\cite{Ott2008cars} for example were performed in motor vehicles under various conditions of moving vehicles with air conditioning (AC) and windows open or closed. 
Additionally, the study by Licina \textit{et. al.}~\cite{Licina2014} was performed to study the human convective boundary layer with particles released by a simulated cough machine.
The very short decay times reported in these experiments are specific to the conditions of the experiments and do not follow the general pattern of indoor contaminant dynamics.

By comparing particle concentrations from direct exposure to the exhalation cloud of an infectious individual and exposure to contaminants in a well-mixed room for different environmental decay rates, we can determine what type of exposure increases the likelihood of infection in a given environment.
The particle dose received directly from an infectious when exposed to their exhalation cloud is given by $d_{cloud}=f_{d}C_{m,I}\lambda_{S}$, where $f_{d}$ is the cone dilution factor for a given distance away from the source, $C_{m,I}$ is the mass distribution of exhaled particles by the infectious, and $\lambda_{S}$ is the breathing rate of the susceptible individual. In this study we take the dilution factor of the exhale cloud \SI{1}{\meter} away from the source to be $f_{d}=0.1$.
The contaminant concentration from a continuous source in a well well-mixed room can be derived by solving the mass conservation equation~\cite{Nazaroff2008}. For a mass that is released into a room of volume $V$ and ventilation rate of $Q$, this concentration is \begin{equation}
    C_{room}(t) = \frac{m}{Q}\left(1-\exp\left(-\frac{Q}{V}t\right)\right),
    \label{eq:room concentration}
\end{equation}
where $Q/V=1/\tau$, and $m$ is the emission rate given in units of mass over time.
If we assume that the emission rate of an infected individual is equal to the breathing rate of a susceptible individual then $C_{m,I}=m/\lambda_{I}=m/\lambda_{S}$ and $m=\lambda_{S}C_{m,I}$, which we can substitute in the relation for $C_{room}(t)$.
The particle dose inhaled by the susceptible in a well-mixed room is then obtained by integrating Eq.~\ref{eq:room concentration} and is $d_{room} = \lambda_{s} \int_{0}^{t} C_{room}(t') dt'$.
For risk calculations in well-mixed environments it is important to consider airborne particles across a wide range of particle sizes~\cite{Pohlker2021, Chong_2021, Wang2021}.

\begin{figure}[!htbp]
	\centering
	\includegraphics[width=0.95\linewidth]{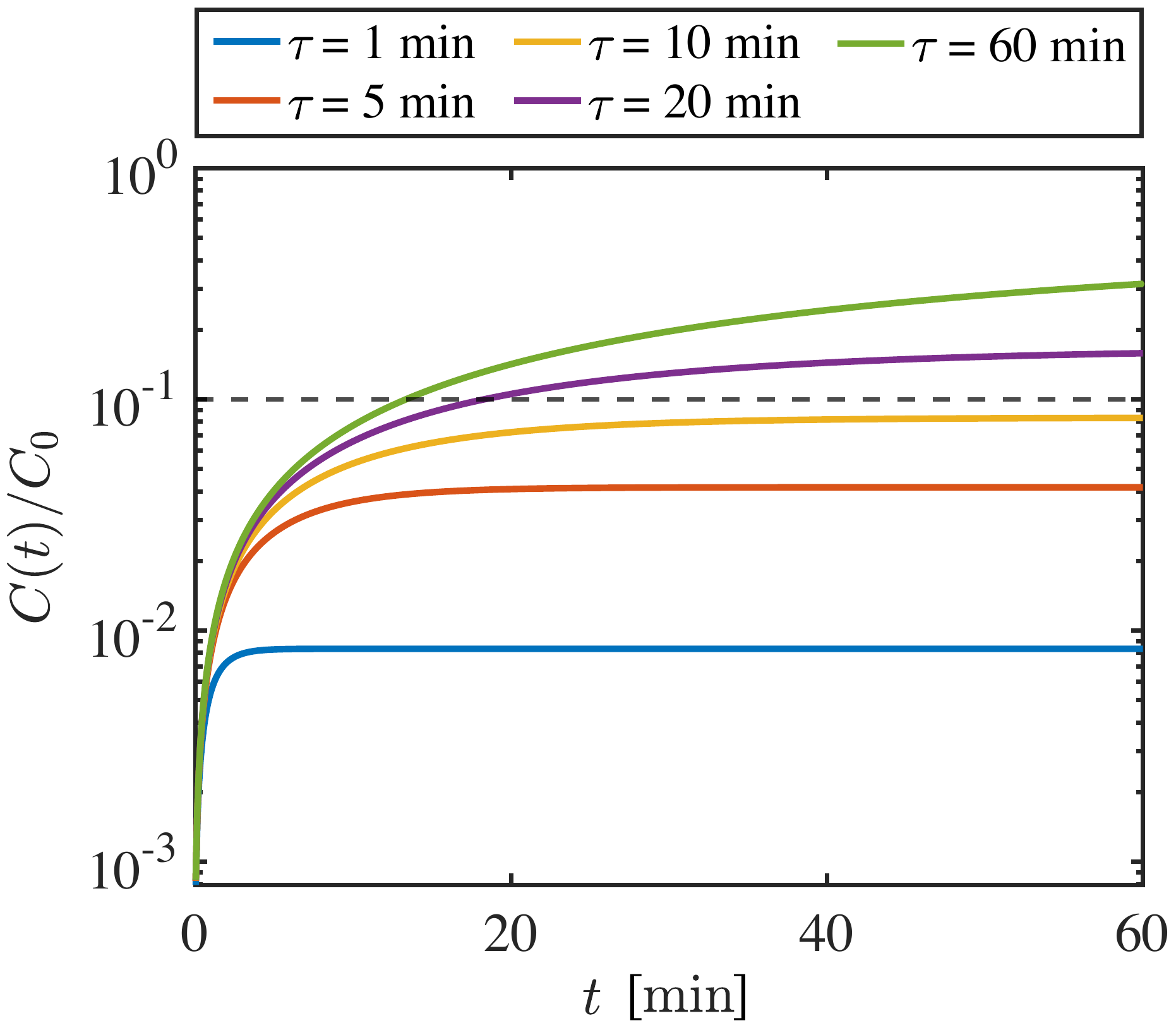}
	\caption{Normalized contaminant concentration as a function of time in a virtual well-mixed box with $V=\SI{1}{\meter^{3}}$ for different decay times $\tau$. The dashed line shows where the contaminant concentration is 10\% of the initial emission $C_{0}$, which is the value we assume for the dilution factor of the exhalation cloud \SI{1}{\meter} away from the infectious.}
	\label{fig:dose ratio}
\end{figure}

To compare exhalation-cloud exposure to the well-mixed room we assume that emissions are concentrated in a virtual \emph{well-mixed box} of volume $\SI{1}{\meter^{3}}$ in the breathing zone of the two subjects, which is a conservative estimate for a distancing scenario in where individuals are \SIrange[range-phrase=-]{1.5}{3.0}{\meter} away from each other.
Fig.~\ref{fig:dose ratio} shows the normalized contaminant concentration of the well-mixed box as a function of time for rooms of different decay times. 
We see that for small $\tau=\SI{1}{\minute}$ the normalized contaminant concentration rapidly reaches a plateau with a concentration that is 10 times smaller than the 10\% threshold expected for the cloud dilution at \SI{1}{\meter}.
Meaning that the contaminant concentration in the exhalation cloud of the infected \SI{1}{\meter} away, is the dominant contributor to infection risk calculations.
These small $\tau$ scenarios are comparable to situations where people interact in calm outdoor environments.
Similarly, for longer $\tau$, the infection risk from contaminants in the well-mixed room is low for \textbf{short} exposure times. 
However as $\tau$ and the exposure time increase, the contaminant concentration also increases and the particles present in a well-mixed environment become more important.
This agrees with our observation that smaller less ventilated rooms pose a higher threat from existing contaminants in the surroundings as compared to larger well ventilated environments with small $\tau$ where the main cause of infection would be direct exposure to the exhalation cloud of infected individuals.

\begin{table}
\fontsize{8}{10}\selectfont
\centering

\begin{tabular}{c c c c}
 \multicolumn{4}{c}{} \\
\hline

Ref. &
Volume ($\text{m}^{3}$) &
ACH ($\text{hr}^{-1}$) &
$\tau$ (min)
\\
\hline

Ishizu\cite{Ishizu1980} & 71 & NA & 7.66\\
\ditto & 268 & NA & 6.83\\
\ditto & 82 & 45.3 & 7.9\\
\ditto & 82 & 8.7 & 8.68\\
 
Leaderer \textit{et. al.}\cite{Leaderer1984} & 34 & 2.3 & 12.9\\

Mage and Ott\cite{Mage1996} & 548 & NA & 2.6\\

Miller and Nazaroff\cite{Miller2001} & 36 & 0.03-1.7 & 23-40\\

Ott \textit{et. al.}\cite{Ott2003house} & 34 & 4 & 44\\

Ott \textit{et. al.}\cite{Ott2008cars} & 3-5 & 3-56.4 & 0.3-7.8\\

Qian \textit{et. al.}\cite{Qian2012} & 242 & NA & 43\\

Stephens \textit{et. al.}\cite{Stephens2013} & 45 & NA & 23.6\\

Licina \textit{et. al.}\cite{Licina2014} & 230 & NA & 0.04\\

Poon \textit{et. al.}\cite{Poon2016} & 11.9 & 40.6 & 11.8\\
 
Hejazi \textit{et. al.}\cite{Hejazi2021} & 10 & 0.0 & 34.3 (13.8-86.5)\\

\ditto & 30 & NA & 10 (7.0-36.2)\\

\ditto & 40,000-150,000 & $<$1 & 1.9 (0.4-15.0)\\

This study & 200 & 0.0 & 171.1 (32.1-358.6)\\

\ditto & 200 & 0.0 & 93.6 (11.1-213.6)\\

\ditto & 200 & NA & 12.4 (7.8-20.4)\\

\ditto & 200 & 6.0 & 9.7 (6.3-11.3)\\

\ditto & 160 & 9.7 & 7.9 (4.8-14.0)\\

\ditto & 170 & 3.2 & 10.9 (7.0-14.0)\\

\ditto & 170 & 4.9 & 7.3 (5.4-8.8)\\

\ditto & 240 & 4.0 & 5.3 (3.0-9.3)\\

\ditto & NA & 4.0 & 9.4 (3.6-11.8)\\

\ditto & 90 & 4.0 & 10.0 (4.8-12.2)\\

\hline

\end{tabular}

\caption{
Experimental observations of indoor contaminant decay rates. All values for $\tau$ measured in this study and Hejazi \textit{et. al.} are average values over $0.3-10 \mu m$ particle size and minimum maximum range in parenthesis.
}
\label{table:lit}
\end{table}

\subsection{Infection risk for COVID-19}
\begin{figure}[htbp]
	\centering
	\includegraphics[width=1\linewidth]{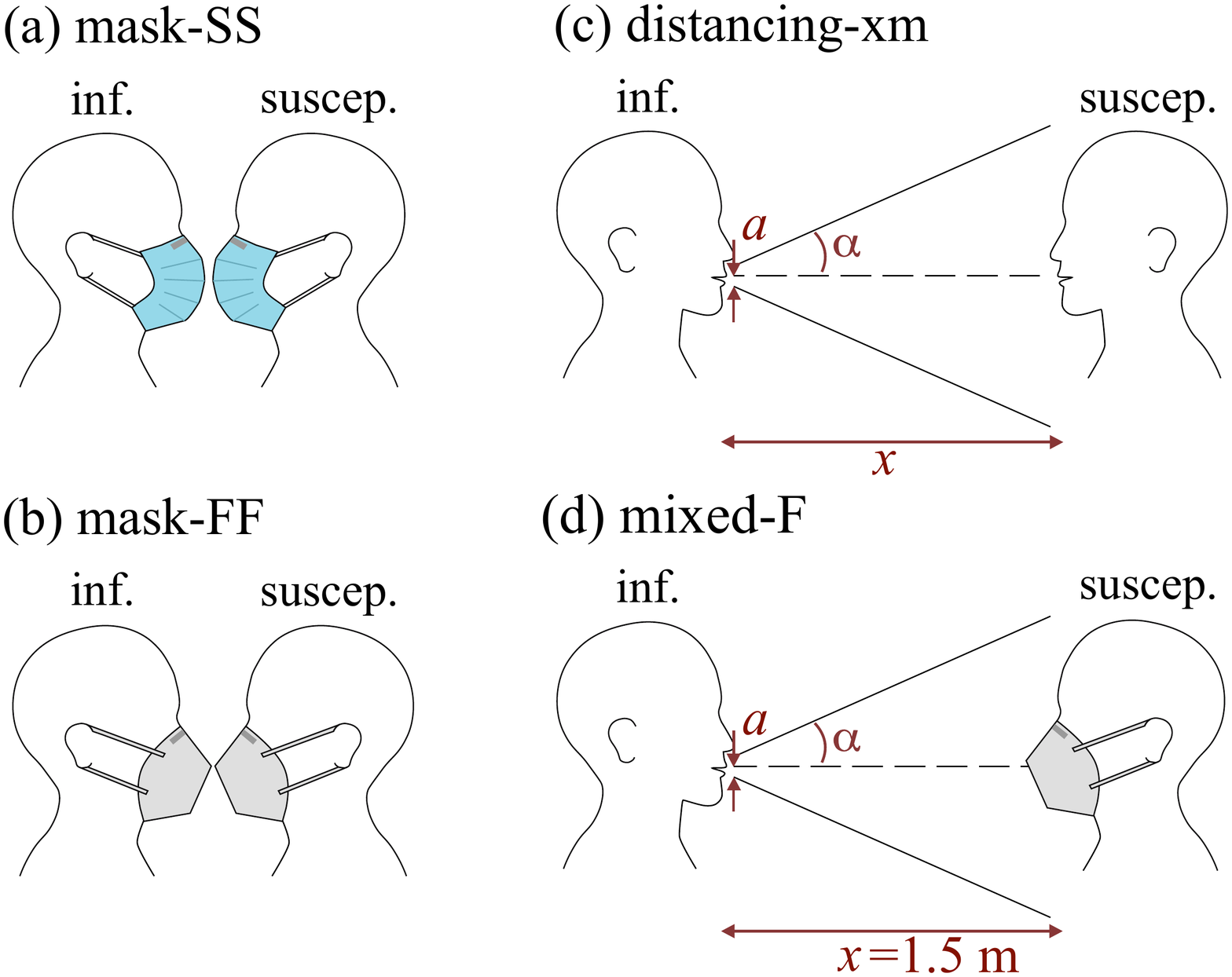}
	\caption{Schematics of scenarios investigated in this study; (a) and (b) \emph{mask-is}: a masked infectious breathing/speaking to a breathing-only masked susceptible, where the susceptible is exposed to the non-diluted total outward leakage of the infectious exhale. \emph{i} and \emph{s} indicate the type of mask worn by the infectious and susceptible individuals, respectively, with FFP2 masks abbreviated by F and surgical masks by S; (c) \emph{distancing-$x$m}: an unmasked breathing-only susceptible exposed to the exhale cloud of an unmasked breathing/speaking infectious while the distance between the two is $x$ meters. In this study $a=$\SI{1.8}{\centi\meter} is the radius of the mouth  and $\alpha=$\SI{10}{\degree} is the exhalation cone half angle; (d) \emph{mixed-s}: the same as (c) but susceptible is wearing a mask and the distance is kept fixed at \SI{1.5}{\meter}. \enquote{mixed-S} indicates the scenario in which the susceptible is wearing a surgical mask and \enquote{mixed-F} indicates an FFP2 mask. }
	\label{fig:scenarios}
\end{figure}

In the following discussions, we assume the FFP2 and surgical masks investigated in our study have characteristics comparable to other FFP2 and surgical masks in terms of total inward and outward leakages and fitting on the subjects' faces. Figure \ref{fig:scenarios} depicts schematics of scenarios considered here, which are:
\begin{itemize}
    \item \emph{Mask} scenario (Fig. \ref{fig:scenarios}a and b): The subjects are in close proximity of each other and we assume that the susceptible is exposed to the total outward leakage of the infectious mask. This provides a safe upper bound for estimating the risk of infection because, in reality, exhaled leakage escapes primarily from below the nosepiece and typically moves upward and away from both individuals. Throughout this text we use \emph{mask-is} nomenclature, where \enquote{i} and \enquote{s} refer to the type of mask used by the infectious and susceptible, respectively, with \enquote{F} denoting FFP2 and \enquote{S} denoting surgical masks. As an example, mask-SF is a mask scenario in which the infectious is wearing a surgical mask and the susceptible is wearing an FFP2 mask.
    \item \emph{Distancing} scenario (Fig. \ref{fig:scenarios}c): the individuals are unmasked and $x$ meters away from each other. We will refer to this as \emph{distancing-$x$m} scenario throughout the text, e.g. distancing-1.5m for a separation of \SI{1.5}{\meter} between two unmasked individuals.
    \item \emph{Mixed} scenario (Fig. \ref{fig:scenarios}d): The same as the distancing scenario but the susceptible is wearing a mask at a distance of \SI{1.5}{\meter} from an unmasked infectious. Throughout the text we use mixed-S and mixed-F to refer to mixed scenarios in which the susceptible is wearing either a surgical or a FFP2 mask, respectively. 
\end{itemize}

\begin{figure}[!htbp]
	\centering
	\includegraphics[width=0.99\linewidth]{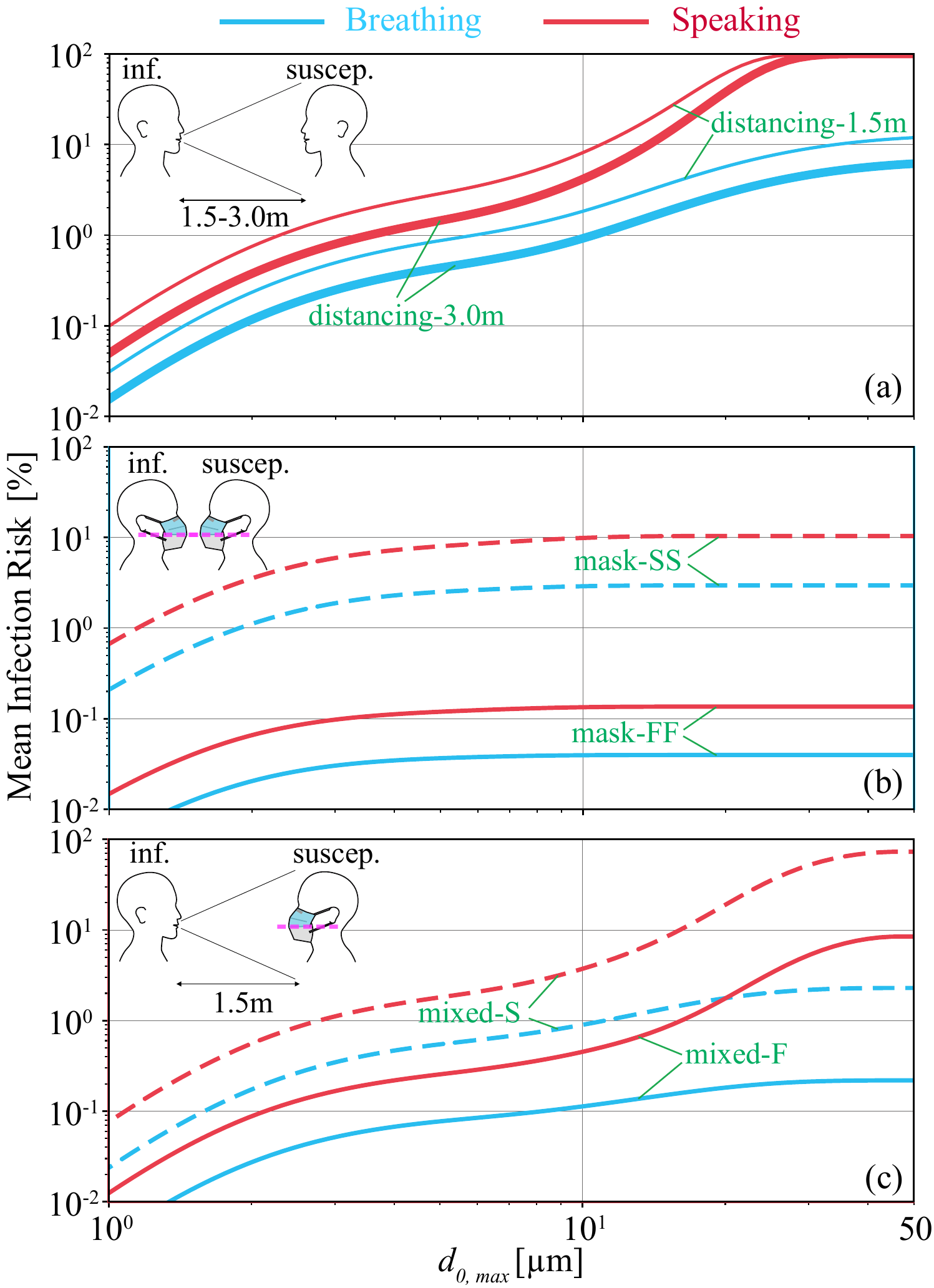}
	\caption{The mean risk of infection as a function of (wet) exhale diameter cut-off $d_{0, max}$ when an infectious is breathing or speaking towards to a breathing-only susceptible for a duration of \SI{20}{\minute} considering (a) distancing, (b) masking and (c) mixed scenarios. Other parameters used for estimating these risks $\rho_p=10^{8.5}\,$\SI{}{virus\,copies\per\milli\liter} and  ID63.21=200.}
	\label{fig:infection_risk_20min}
\end{figure}

Fig. \ref{fig:infection_risk_20min} shows the mean risk of infection for the different scenarios and for a duration of \SI{20}{\minute} as a function of wet diameter cut-off (the diameter above which the particles are assumed to deposit quickly and do not contribute to airborne infection risk). 
The first observation that can be made is that if a \SI{5}{\micro\meter} cut-off is considered, the typical cut-off size for \emph{aerosols} \cite{Pohlker2021, randall2021did}, risk of infection is below 10\% for all the scenarios. 
However, at cut-off size of \SI{50}{\micro\meter}, which in typical room conditions with $w=4$ translates to an equilibrium diameter of $ 12.5\,$\SI{}{\micro\meter}, risk of infection increases significantly for distancing and mixed scenarios, particularly when infectious is speaking. 
While the trend of mean infections risk for breathing and speaking infectious is very similar for a cut-off size of \SI{10}{\micro\meter}, significant deviations for larger cut-off sizes are visible in \ref{fig:infection_risk_20min} (a) and (b). 
The reason for such a significant deviation between breathing and speaking is the increased probability of producing \textgreater\SI{10}{\micro\meter} for vocalization-associated activities. However, this deviation in the trend disappears for masking scenario, \ref{fig:infection_risk_20min} (b), in which the \SI{10}{\micro\meter} cut-off is enforced by the masks themselves, as shown in Figures \ref{fig:maskleakage} and \ref{fig:effective_penetration}.

The distance traveled and residence time of even \textgreater\SI{50}{\micro\meter} particles in the air have recently been shown to be much larger than previously thought \cite{Chong_2021, Wang2021}. 
Thus, the mean infection risk shown in Fig. \ref{fig:infection_risk_20min} are very plausible upper limits for a \SI{50}{\micro\meter} cut-off and we restrict ourselves below to the results obtained for this cut-off. 

Fig. \ref{fig:infection_risk_20min} also shows that increasing distancing from \SI{1.5}{\meter} to \SI{3.0}{\meter} reduces risk of infection when infectious is breathing but not for a speaking infectious.  
In contrast, \emph{collective masking} is an extremely effective strategy in reducing risk of infection, even for the absolutely extravagant scenario considered here, in which the susceptible inhales all the infectious' exhaled air with the leakage from an adjusted mask is also accounted for (other mask type/leakage combinations are discussed below).
Finally, Fig. \ref{fig:infection_risk_20min}(c) shows that when only the susceptible adheres to masking and even when they are distancing, the probability of infection risk can be as high as ${\sim}10\%/70\%$ for being in the exhalation cloud of a breathing/speaking infectious.
Fig. \ref{fig:infection_risk_20min} (b) and (c) show that an adjusted FFP2 mask reduces risk of infection by a factor of 10 compared to an adjusted surgical mask. 
When both infectious and susceptible wear FFP2 masks, i.e. Fig. \ref{fig:infection_risk_20min} (b), a reduction in risk of infection by a factor of $\sim 80$ is expected when they both wear surgical masks. 

\begin{figure}[!htbp]
	\centering
	\includegraphics[width=0.99\linewidth]{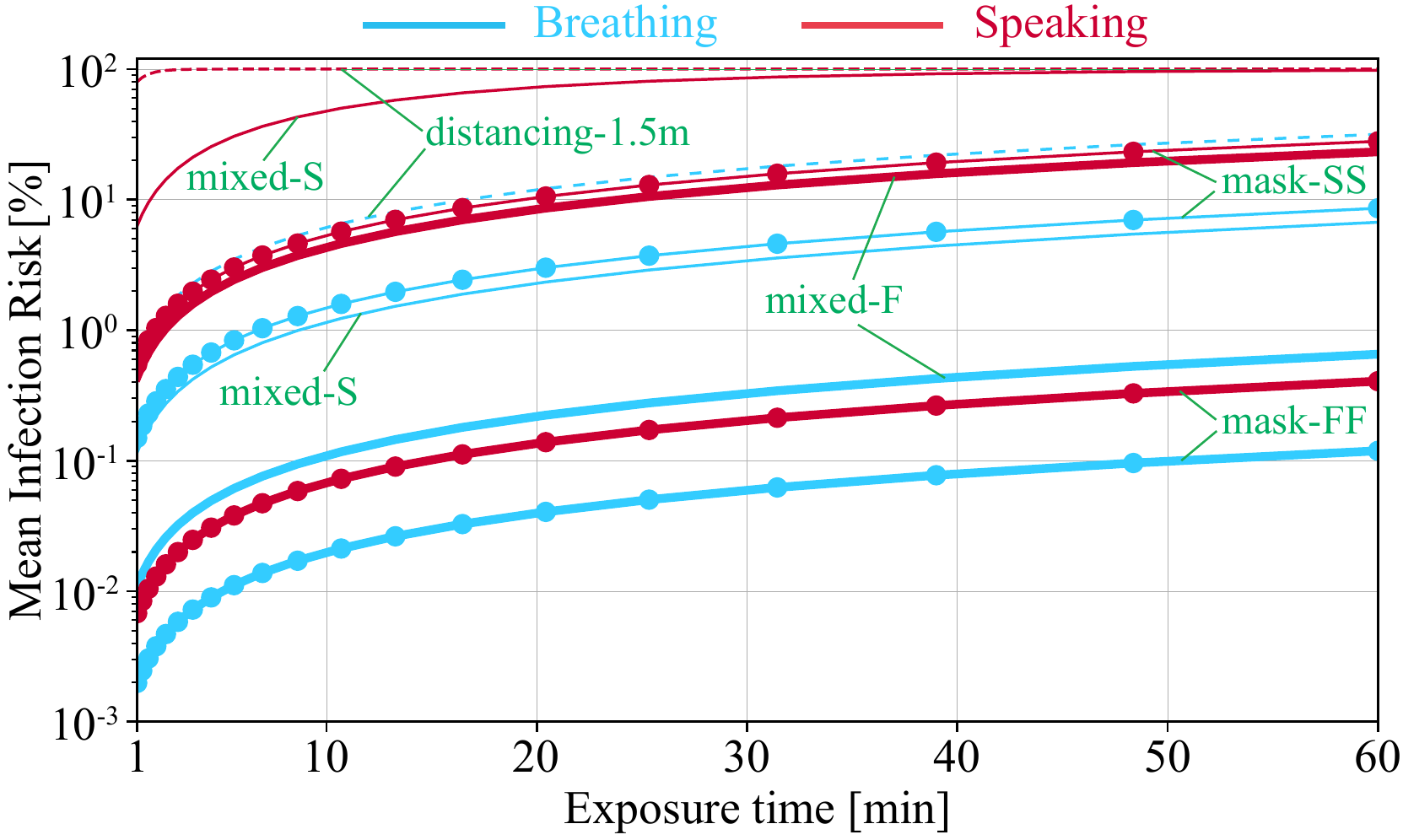}
	\caption{ Mean risk of infection for a breathing-only susceptible to be exposed to a breathing or speaking infectious in different scenarios as a function of time. Other parameters used for estimating these risks $\rho_p=10^{8.5}\,$\SI{}{virus\,copies\per\milli\liter} and  ID63.21=200.}
	\label{fig:infection_risk_time}
\end{figure}

Fig. \ref{fig:infection_risk_time} shows risk of infection as a function of time for different scenarios.
Let us consider an infection risk of 1\% as the threshold beyond which a given scenario is unsafe.
Given this threshold, the distancing scenarios quickly becomes unsafe and already after about\SI{1.5}{\minute} for a speaking infectious, the risk of infection for the susceptible at a distance of \SI{1.5}{\meter} is 90\%. 
The next high-risk scenario is the mixed-S scenario with a speaking infectious, which surpasses the 1\% threshold in less than a minute and reaches the 90\% threshold after half an hour. 
All the speaking infectious scenarios with the exception of mask-FF bypass the 1\% infection risk threshold within a few minutes and reach \textgreater 10\% in one hour. 
The only breathing infectious scenario associated with \textgreater 10\% infection risk in one hour is the distancing-1.5m.  
The safest scenarios that stay below the 1\% threshold for one hour of exposure in order of best to worse are mask-FF for breathing and speaking infectious, respectively, followed by the mixed-F with a breathing infectious.
Interestingly, the extremely conservative estimate of mask-FF with speaking infectious is safer than mixed-F with a breathing infectious. 
This is yet another indicator showing the effectiveness of collective masking. 

\begin{figure}[!htbp]
	\centering
	\includegraphics[width=0.99\linewidth]{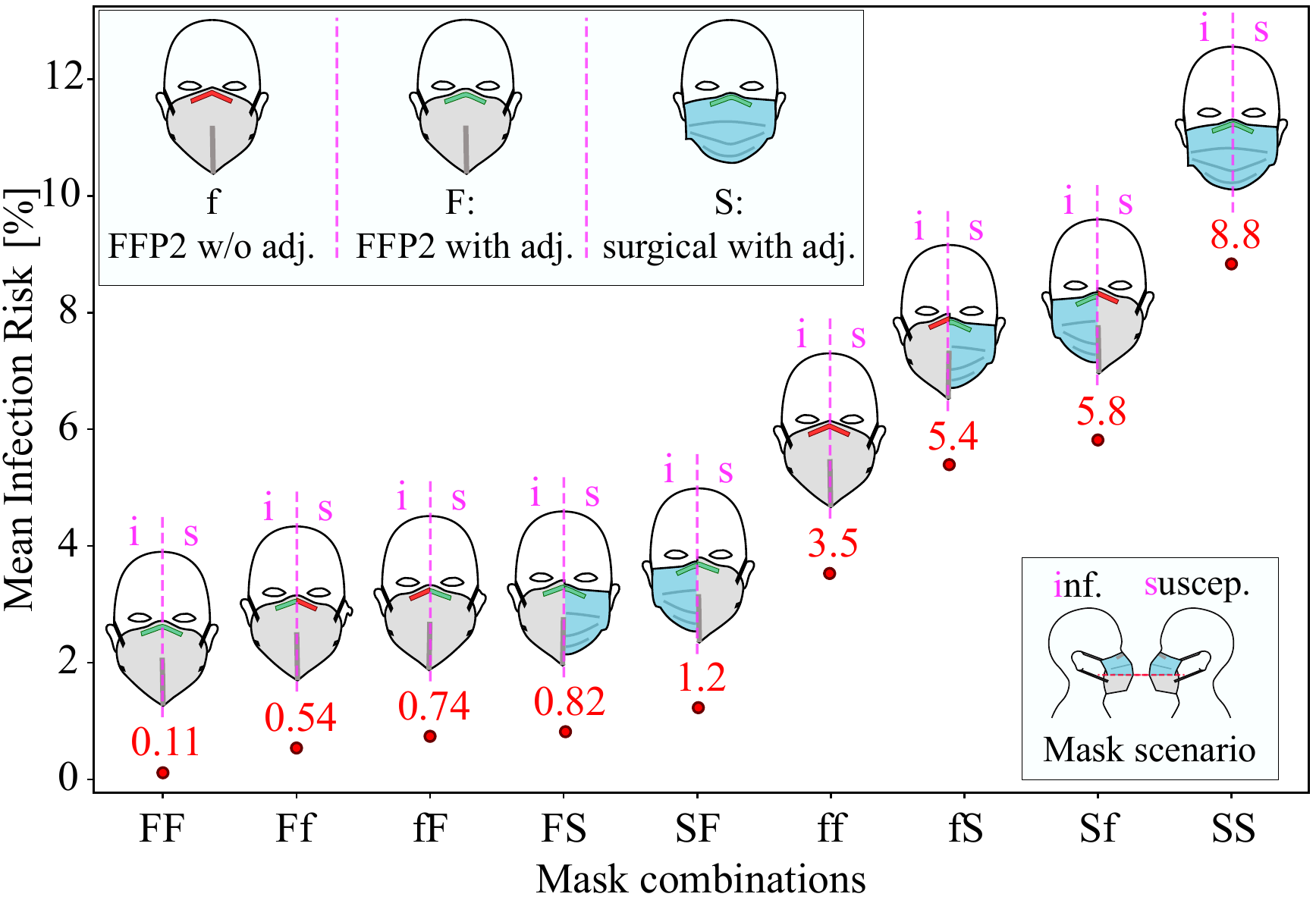}
	\caption{Mean risk of infection for masked susceptible to be next to a speaking and masked infectious for \SI{20}{\minute}, i.e. speaking mask-is scenario where \enquote{i} indicate the type of mask worn by the infectious and \enquote{s} the type of mask worn be susceptible. Mask types and fittings are abbreviated as follow: f: FFP2 mask without adjustment (Fig. \ref{fig:maskleakage} case i), F: FFP2 mask with adjustment (Fig. \ref{fig:maskleakage} case ii), S:  surgical mask with adjustment (Fig. \ref{fig:maskleakage} case v). Other parameters used for estimating these risks $\rho_p=10^{8.5}\,$\SI{}{virus\,copies\per\milli\liter} and  ID63.21=200.}
	\label{fig:infection_risk_mask_combo}
\end{figure}

Fig. \ref{fig:infection_risk_mask_combo} shows different combination of FFP2 fittings (with and without nosepiece adjustments) and nosepiece-adjusted surgical masks for a speaking infectious considering \emph{mask-is} scenario and an exposure duration of \SI{20}{\minute}. 
The best preventive measure is obviously adjusted FFP2 masks (case FF) for both infectious and susceptible, and the least-safe one is surgical masks (case SS) for them. 
Interestingly, very loosely fitted FFP2 masks (case ff) outperform adjusted surgical masks (case SS) by a factor of 2.5.
Proper nosepiece adjustment for FFP2 masks can decrease risk of infection by a factor of 30 (case FF vs case ff), while if at least one of the infectious or susceptible adjust their FFP2 masks the increase in risk compared to case FF is about 5-7 times. 
Risk of infection for asymmetric cases, i.e. Ff vs fF, FS vs SF and fS vs Sf, is lower by about 7\%-50\% when the better mask or the better adjusted mask is worn by the infectious. 

We have also investigated whether there is any significant difference between the poly-pathogen and mono-pathogen models, and found that for the scenarios considered here the differences are very negligible.  
It is expected in scenarios with higher viral loads and for activities associated with production of larger particles, e.g. sneezing, the differences become more significant. 
Furthermore, we have found that changing shrinkage ratios from 1 to 4 would double the infection risk for mask scenario and \textless30\% for distancing scenario.

\section{Conclusion}
We have shown that during short exposure periods, direct exposure, or in large indoor spaces, even conservative estimates of well-mixed room models significantly underestimate infection risk. In such environments, the near-field model, in which the susceptible is in the exhalation cloud of the infectious or wearing masks by having the masked susceptible inhale the entire exhalation of a masked infectious, can provide a reliable upper bound of the risk of infection from airborne disease. Our results show that social distancing alone without masking is associated with a very high risk of infection, especially in environments where individuals are speaking. High infection risks are also expected when only the susceptible wears a mask, even with social distancing. We show that collective masking is the most effective method for limiting airborne transmission of SARS-CoV-2, even when faceseal leaks are considered.  The main factor affecting infection risk in the collective masking scenario is leakage between the mask and the face. The fitted FFP2 masks studied here (and most likely other vertically folded FFP2 masks of similar design), when properly fitted to infectious and susceptible faces, can reduce the risk of infection by a factor of 30 compared with loosely worn masks and by a factor of 80 compared with fitted surgical masks.  Our results also suggest that the use of FFP2 masks should  be preferred to surgical masks, as even loosely worn FFP2 masks can reduce the risk of infection by a factor of 2.5 compared with well-fitted surgical masks. For infectious and susceptible, the better mask/fitting should be preferred for infectious to achieve a lower risk of infection.  Nonetheless, we found that both surgical masks and FFP2 masks are very effective in minimizing infection risk, especially considering that the upper bound on infection risk used here is intentionally extremely conservative.

\subsection{Methods}
\subsection{Mask efficacy measurements}
\label{sec: mask_efficacy}
\subsubsection{Filter penetration}
\label{sec:penetration_methods}
Two different setups were used to measure the particle size-dependent penetration through the filter material of different masks. In the first setup, hereafter referred to as \enquote{setup 1}, three cloth, seven surgical, one FFP1 and four FFP2 masks were tested. Each mask was fixed between two aluminum plates which were pressed together in a bench vice. An o-ring was used to ensure a good seal between the two metal plates and the mask.
Both metal plates have a circular opening. The air can enter the mask from one side. The other side is connected to the sampling tube which leads via a Y-connection to the Optical Particle Sizer 3330 spectrometer (OPS) and the NanoScan Scanning Mobility Particle Sizer 3910 (NS-SMPS) from TSI. With this 
the filtered particle concentration $c_{filter}$ was measured. With the combination of OPS and NS-SMPS, particles in the size range of \SI{10}{\nano \meter} to \SI{10}{\micro \meter} were counted in $13+16$ logarithmically spaced bins. Since the OPS spectrometer has a sampling flow rate of \SI{1}{\liter\per\minute} and the NanoScan NS-SMPS of \SI{0.76}{\liter \per \minute}, the total sampling flow rate was \SI{1.76}{\liter\per\minute} through the circular area with diameter of \SI{3.2}{\centi\meter}. Applied to the whole penetrable mask surface of for example one FFP2 mask the flow velocities through the mask fabric in setup 1 (\SI{3.7}{\centi \meter \per \second}) are equivalent to those of a total flow rate of \SI{51}{\liter\per\minute} through the whole mask (similar for other masks as mask surface areas are comparable). The particles found naturally in the outdoor air were used as test particles. The background concentration $c_{bg}$ was measured with the same combination of OPS and NS-SMPS before and after each filter measurement. Both background and filter measurements were performed over at least \SI{3}{\minute} each. The arithmetic mean of before  and after measurements was used as the background for the penetration calculation. Please note that  naturally occurring particles in outdoor air were found to be almost  constant over time.
The last bin of the NS-SMPS was removed from the analysis as it overlaps with the first OPS bin which has better counting statistics by design. Particle diameters measured with the NS-SMPS (below \SI{0.3}{\micro \meter}) are electron mobility diameters whereas the OPS measures the optical diameter.

The second setup, hereafter referred to as \enquote{setup 2}, was used to measure particle penetration through the filter material of masks used in the leakage experiments in this study (one type surgical and one type FFP2 mask, the masks' dimensions and designs are presented in the supplementary information, Fig. S2). Setup 2 is similar to setup 1 but the diameter of the circular opening in the metal plate was reduced to \SI{1.6}{\centi\meter}. With this smaller opening we have a flow velocity of (\SI{14.6}{\centi \meter \per \second}) through the mask, equivalent to a total flow rate of \SI{204}{\liter\per\minute} through the whole FFP2 mask and \SI{188}{\liter \per \minute} through the whole surgical mask (which exceeds the \SI{95}{\liter\per\minute} required for FFP testing by the European standard). For setup 2, dolomite dust from DMT GmbH \& Co KG that has particle size distribution of \textless \SI{20}{\micro \meter} was used as background particles in a well mixed closed room. For each mask, the filtered particle concentration was measured for 1 minute and the background was measured for 1 minute before and after each measurement. Since the background particle concentration exponentially decays, the geometric mean of the before and after background samples was used as the background concentration for the filter penetration calculation,
where the filter penetration is defined as $P_{filter}=c_{filter}/c_{bg}$.

\subsubsection{Total inward leakage}
\label{sec:til_methods}
The total inward leakage $TIL$ of a mask or filtering facepiece is defined as the ratio of particle concentration inside the mask $c_{mask}$ to particle concentration in the background $c_{bg}$, i.e. $TIL=c_{mask}/c_{bg}$.
The $TIL$ is also equivalent to the sum of the inward filter penetration $P_{in}=P_{filter}$ and inward particle penetration through face seal leaks $L_{in}$, hence, $TIL=P_{in}+L_{in}$.
To measure particle size-dependent total inward leakage from the surgical and FFP2 masks with different fittings, worn separately or together, we have investigated several combinations of masks/fittings (also see legend Fig. \ref{fig:maskleakage}):

\begin{enumerate}[label=(\roman*),topsep=-1ex,itemsep=-1ex,partopsep=1ex,parsep=1ex]
\item FFP2 mask without fitting the nosepiece (also called nose clip or nose wire) masks are usually shipped with a sharp bend
\item vertical-fold FFP2 mask detailed fitted to the face by reshaping the nosepiece and
adjusting the mask so that it fits closer to the subjects face,
\item FFP2 mask with a surgical mask over it, both with adjustment of the nosepiece fitting by pre-bending,
\item FFP2 mask with a 1 by \SI{~12}{\centi\meter} double-sided 3M\textsuperscript{TM} Medical Tape 1509 applied underneath the nosepiece to fully seal the mask to the nose and areas around it,
\item surgical mask with adjustments to the nosepiece fitting.
\end{enumerate}

These cases, have been measured on seven adult subjects (one female and six males, three of whom had noticeable facial hair; subjects' facial dimensions are presented in the supplementary information, Tab. S1) while breathing normally through the nose.
The experiments were performed in a \SI{200}{\cubic \meter} room. Before starting measurements the dolomite dust was released into the room by shaking a dust-covered micro-fiber cloth in front of a \SI{120}{\watt} fan with \SI{0.26}{\meter} blades to mix the particles in the room air.
The fan was running throughout the measurements while it was \SI{3}{\meter} away from the subject and oriented at an angle of \SI{\sim30}{\degree} towards the ceiling to reduce potential non-isokinetic sampling bias in the measurement of background concentrations. 
New particles were released into the room periodically to compensate for the loss of large particles in the background due to deposition on the floor and fan blades. The total inward leakage of the different cases (i) to (v) were examined for each subject while they were seated in a chair. Subjects were in a relaxed sitting position for at least one minute before measurements began. 
Two OPS spectrometers (cf.\ methods~\ref{sec:penetration_methods}) were used synchronously to measure the background and inhaled air samples simultaneously. 
A sampling resolution of \SI{1}{\second} was chosen and total inward leakage was measured for a duration of at least \SI{100}{\second} for each case (a total duration of \SI{\sim15}{\minute} per subject). 
The sampling flow rate of both spectrometers was \SI{1}{\liter \per \minute}. The sampling tube measuring in-mask concentrations $c_{mask}$ was held in place by an easily adjustable arm attached to a helmet worn by the subjects as shown in the supplementary information (SI, Fig. S3). 
The in-mask sampling tube was connected to a plastic feed-through that passed through a punched hole in the mask with a diameter of \SI{8}{\milli\meter} and was tightened onto the mask from the inside with a nut. 
The horizontal position of the feedthrough was \SI{1}{\centi\meter} from the middle seam of the mask and the vertical position was chosen for each subject individually so that it was half-way between the nose and upper lip. Test experiments with different locations  showed that the chosen inlet position did not lead to significantly different results than a lower inlet position between the lower lip and chin (see SI, Fig. S6).
The helmet arm for holding the inhalation sampling tube was adjusted for each subject so that the tube neither pressed on nor pulled on the mask. Moreover, the last \SI{35}{\centi \meter} of the sampling tube were made out of a more flexible rubber material compared to the conductive PTFE tube to minimize any forcing on the mask. The background concentration $c_{bg}$ was measured approximately \SI{20}{\centi \meter} in front of the subjects head.
Both sampling tubes (in-mask and background) had the exact same lengths and approximately the same curvatures. In any case, possible differences between the particle paths and OPSs were corrected by performing sensitivity corrections as described in SI 1D1. Furthermore, each sampling tube was connected to a diffusion dryer (TOPAS DDU570/H) to remove humidity that could alter measured particle sizes (see SI, Fig. S5 for comparison against measurements without diffusion dryer). All tubes and connections were checked for potential leaks before measurements by using a High-Efficiency Particulate Arrestance (HEPA) filter and observing zero particle counts on the spectrometers. 

\subsubsection{Mask data analysis}
To correct for varying sensitivities in the different bins of the spectrometers (OPS and NS-SMPS) a running geometric average over 3 bins each was used. For the leakage measurements that were performed with two OPS spectrometers simultaneously, the possibly varying sensitivities and different particle loss rate inside the tubes and diffusion dryers were corrected with a 27-minute long simultaneous calibration background measurement.
For the total inward leakage calculation, only the in-mask particle concentration during should be included in the analysis \cite{myers1986parameters}. Inhalation is associated with a peak in in-mask particle concentration. Therefore, only samples within the top 10\% of a peak in total particle count were considered for the total inward leakage analysis. More details on this can be found in the supplementary information (SI, section 1D2).

\subsubsection{Total outward leakage}
\label{sec:TOL}
The total outward leakage $TOL$ is the sum of outward filter penetration $P_{ex}$ and the outward faceseal leakage during exhalation $L_{ex}$, i.e. $TOL=P_{ex}+L_{ex}$.
Filter penetration should not depend on flow direction and therefore $P_{ex}=P_{in}=P_{filter}$.
Outward leakage of masks was not measured in this study. Previous studies on outward leakage paint an inconclusive picture.
Van der Sande \textit{et al.}\ \cite{van2008professional} found larger total outward leakage (smaller outward protection factors) for an FFP2 and surgical mask compared to total inward leakage. The difference is more significant for the FFP2 mask. This could be explained by a high pressure difference between inside and outside the mask which pushes the mask away from the face leading to higher leakage \cite{mittal2020flow}. However, in their study total inward leakage was measured on human subjects, whereas total outward leakage was studied on a manikin which could influence mask fit.
In contrast, a pure manikin study by Koh \textit{et al.}\ \cite{koh2021outward} found no difference between inward and outward leakage for a fitted N95 respirator. For an unfitted N95 mask and a surgical mask, outward leakage was measured to be slightly lower than inward leakage, which was also observed in another manikin study by Pan \textit{et al.}\ \cite{pan2021inward} for a surgical mask. In ref. \cite{pan2021inward} the difference between inward and outward leakage was found to vary from mask to mask. For the surgical mask inward leakage was also larger than outward leakage.
These inconclusive results combined with uncertainties in the experimental methods lead us to assume $L_{in}\approx L_{ex}$ and therefore $TOL \approx TIL$ for all measured mask cases in our study.

\subsection{Infection risk model}
Considering the fact that exhaled particles can contain one or more pathogens, depending on their size and pathogen concentration, the average infection probability, hereafter referred to as infection risk, can be calculated via \cite{Nordsiek2021}:

\begin{equation}
	R_I =  1 - \exp{\left[-\sum_{k=1}^\infty{\left(1 - (1 - r)^k\right) \mu_k}\right]} \quad ,
	\label{eqn:modified_exponential_model}
\end{equation}

\noindent where $\mu_k$ is the absorbed \emph{aerosol dose} with multiplicity of $k$, i.e. containing k pathogens inside and $r$ is the probability of each pathogen to cause infection. Note that $r = 1/D$, where $D$ is the infectious dose required for $63.21$\% chance of infection or ID\textsubscript{63.21}. Details on how to calculate the absorbed \emph{aerosol dose} are given in \citep{Nordsiek2021}. In the traditional exponential model only particles with $k=1$ are considered, in which case Eq. \ref{eqn:modified_exponential_model} reduces to $R_{I, mono} = 1 - e^{-r\, \mu_1}$, where $\mu_1$ is the average number of mono-pathogen-borne particles absorbed. The mono-pathogen infection risk $R_{I, mono}$ always overestimates the infection risk compared to the poly-pathogen $R_I$ formulation shown in Eq. \ref{eqn:modified_exponential_model} \cite[see][for more details]{Nordsiek2021}. In order to calculate the infection risk and the average absorbed dose, the following assumptions are made:

\begin{itemize}[topsep=-1ex,itemsep=-1ex,partopsep=1ex,parsep=1ex]
    \item Initial pathogen concentration in the room and initial absorbed pathogen dose of the susceptible are both zero.
    \item Airborne transmission from only one infectious to only one susceptible is considered. 
    \item Susceptible is always in the exhale cloud of the infectious. 
    \item Pathogen accumulation in the ambient is negligible, thus, susceptible can inhale pathogens only when the infectious is active. As a result, exposure duration is smaller than or equal to source-active duration. The validity of this assumption is going to be discussed later. 
    \item Exposure duration is much shorter than the time needed for the pathogen inactivation to be significant. 
\end{itemize}

Given these assumptions, the total absorbed dose by the susceptible individual can be calculated as follows:

\begin{equation}
\begin{aligned}
 \mu_{k}(t)   = \int_{d_{0, min}(k)}^{d_{0, max}(k)}  \, \mathrm{d}\phi \int_{0}^{t_{exp}} \mathrm{d}t &\!  \overbrace{ n_{I, k}(\phi, t) \,  f_d\left(\phi, \lambda_{I}(t), w(\phi, t), t\right)}^\text{infec. particle conc. in breath. zone of susceptible indv.} \\
	& \times  \overbrace{\left[ P_{ex}(\phi, \lambda_{I}(t))+L_{ex}(\phi, \lambda_{I}(t))\right]}^\text{total outward leakage}\\
	& \times  \overbrace{\left[ P_{in}(\phi, w(\phi, t), \lambda_{S}(t)) +L_{in}(\phi, w(\phi, t), \lambda_{S}(t))\right]}^\text{total inward leakage}\\
	& \times  \overbrace{D_{rt}(\phi, w(\phi, t), \lambda_{S}(t))}^\text{deposition in respiratory tract}\, \overbrace{\lambda_{S}(t)}^\text{inhaled rate} \quad ,
\end{aligned}
	\label{eqn:absorbed_dose_full}
\end{equation}

\noindent where 
\begin{equation}
	n_{I, k}(d_0,t) =
	\begin{cases}
		C_{n, I}\left(d_0, t\right) e^{-\expectedk{d_0}} \left(\frac{\expectedk{d_0}^k}{k!}\right) & \text{if $d_0 \ge d_{0, min}(k)$} \quad , \\
		0 & \text{if $d_0 < d_{0, min}(k)$} \quad ,
	\end{cases}
	\label{eqn:infectious_production_nk_by_d0}
\end{equation}

\noindent and $\expectedk{d_0} = \frac{\pi}{6} d_0^3 \rho_{p}$, $d_0$ is the initial wet particle size on exhalation by the infectious, $w(d_0, t) = d_0/d_e$ is the shrinkage factor defined as the ratio of the particle initial wet diameter $d_0$ to the equilibrium diameter $d_e$ after it is exposed to the typically sub-saturated conditions of the room and lost its volatile components, $d_{0, min}(k)$ and $d_{0, max}(k)$ are the minimum and maximum particle size that can be aerosolized and contain $k$ copies of the pathogen, $t_{exp}$ is the exposure duration of the susceptible individual, $\rho_{p}$ is the pathogen number concentration, i.e. the viral load, in the infectious respiratory-tract-fluid, $C_{n,I}(d_0, t)$ is the number concentration of exhaled particles at the mouth/nose of the infectious, $f_d(d_0, \lambda_{I}(t), w(d_0, t), t)$ is the fractional ratio at which the particle concentration of the exhaled air by the infectious individual decreases until it reaches the breathing zone of the susceptible individual due to (turbulent and/or molecular) mixing with the room air or particle deposition losses, $P_{ex}(d_0, \lambda_I)$ is the outward filter penetration of the face-mask fabric worn by the infectious, $L_{ex}(d_0, \lambda_I)$ is the outward faceseal leakage of the face mask worn by the infected individual during exhalation, $P_{in}(d_0, w(d_0, t), \lambda_S)$ is the inward filter penetration of the face-mask fabric worn by the susceptible, $L_{in}(d_0, w(d_0, t), \lambda_S)$ is the inward faceseal leakage of the face mask worn by the susceptible, $D_{rt}(d_0, w(d_0, t), \lambda_{S}(t))$ is the deposition efficiency of the inhaled particles within the respiratory of the susceptible individual, and $\lambda_{I}(t)$ and $\lambda_{S}(t)$ are the volumetric inhalation rate (also called ventilation rate) of the infectious and susceptible, respectively. It should be noted that many of the parameters present in Eq. \ref{eqn:absorbed_dose_full} are also functions of the room conditions, e.g. relative humidity, temperature, ventilation type, air velocity, which are neglected here.

We assume that the $\rho_{p}$ is constant and independent of the particle size, even though it has been shown that particles of different sizes have different production sites within the respiratory tract \cite{Pohlker2021} and particles of different origins might have different viral loads \cite{hou2020sars}. The SARS-CoV-2 viral load, $\rho_p$, is in the very broad range of 10\textsuperscript{2}-10\textsuperscript{11}\SI{}{\per\milli\liter} \cite{Miller2021}. Mean values for the currently measured SARS-CoV-2 variants are 10\textsuperscript{8.2}-10\textsuperscript{8.5}\SI{}{\per\milli\liter} \cite{kissler2021densely}. Here we use 10\textsuperscript{8.5}\SI{}{\per\milli\liter} to obtain an upper estimate on risk of infection, which should be more applicable to the new variants of the SARS-CoV-2. The increase in viral load with the new variants currently circulating globally is constant with findings in other studies \cite[e.g. see][and references therein]{Jones2021}.  $C_{n,I}$ values are obtained from the database in \cite{heads}, which is collected based on measurements from more than 130 subjects aged 5-88 years using aerosol size spectrometers and in-line holography covering wet particle sizes, i.e. $d_0$, from \SI{10}{\nano\meter} up to \SI{1}{\milli\meter}. The smallest particle size considered for infection risk analyses, i.e. $d_{0, min}$, is \SI{0.2}{\micro\meter}, which is about two times the size of the SARS-CoV-2 virus \cite[e.g. see][]{Pohlker2021, Nordsiek2021}. As for the upper limit, we considered $d_{0, max} = \SI{50}{\micro\meter}$ and assumed larger particles typically deposit to the ground very quickly and in the vicinity of the infectious person. However, it should be noted that there is an ongoing debate regarding the advection distance of exhaled particles in different respiratory activities and room conditions \cite[e.g. see][and references therein for more details]{Chong_2021, Wang2021}.

Particles exhaled by the infectious are moist and, depending on the relative humidity, may decrease considerably in size by evaporation until they reach the breathing zone of the susceptible. Unless otherwise stated, we have assumed all the particles shrink by a factor of 4, i.e. $w = 4.0$, which is the expected shrinkage factor for relative humidity encountered in typical indoor environments \cite{Pohlker2021}. For a \SI{50}{\micro\meter} water droplet with 0.005 volumetric fraction of NaCl, which can be considered as a very simplified surrogate to human saliva, it takes about 4-10 seconds (1-1.7 seconds for a \SI{20}{\micro\meter} NaCl-water droplet) to reach equilibrium diameter in stagnant air with relative humidity of 40-70\%. As a result, we assume the average age of inhaled air is longer than a few seconds. Nonetheless, given that this cannot be assumed to be always the case, especially in near field exposures, the sensitivity of the results from our simulations on $w$ are also investigated. 

The values published in Table 15 of \cite{ICRP1994} are used to calculate $\lambda_{I}(t)$ and $\lambda_{S}(t)$.
However, since these rates are given for general physical activities, i.e., sleeping, sitting, and light and heavy exercise, they are combined by optimal weighting factors that were found iteratively and that reproduced the rates found in the literature for different respiratory activities \cite{hoit1990speech, Gupta_2009, Gupta_2010}. Values of breathing and speaking ventilation rates assumed to be constant and equal to \SI{0.57}{\cubic\meter\per\hour} and \SI{0.67}{\cubic\meter\per\hour}, respectively.

While $P_{ex}$ and $L_{ex}$ are functions of particles diameter during exhalation; $d_0$, $P_{in}$, and $L_{in}$ are dependent on particle diameter during inhalation, $d_e = d_0/w$. The penetration of mask fabric is also a function of breathing rates since it will influence the particle loss due to inertial impaction (important for supermicron particles) and the time required for capturing submicron particles due to Brownian diffusion. The penetration due to mask leakage is also a function of particle diameter and breathing rate, more details regarding these parameters can be found in methods section~\ref{sec: mask_efficacy}. 

The ICRP respiratory tract deposition (ICRP94) model \cite{ICRP1994} is used to calculate $D_{rt}(d_0, w(d_0, t), \lambda_{S}(t))$,
The ICRP94 model can provide an estimate of particle deposition in five different regions of the respiratory tract based on empirical and numerical models, namely nasal, oral, thoracic bronchial, bronchioles, and alveolar regions.
In order to capture the deposition of exhaled particles that have dried in the typically sub-saturated air of a room, one also needs to consider that such particles will undergo hygroscopic growth as they enter the almost saturated environment  within the respiratory tract, i.e. with a relative humidity of 99.5\% \cite{Ferron1990, ICRP1994, Broday2001, Pohlker2021}. To take into account the hygroscopic growth of inhaled particles, the coupled equations describing rate of change in the particle size and its temperature are solved simultaneously,  as explained well in section 13.2.1 of \cite{pruppacher2010microstructure}, assuming fully dried particles consisting of  pure NaCl crystals. This assumption is a good approximation for human aerosols, although a more detailed knowledge would be highly beneficial. The osmotic coefficient required for hygroscopic growth of the NaCl solution is calculated via formulations provided by \cite{TANG1986409}. The hygroscopic growth codes are verified against diffusional growth rate curves shown in Fig. 13.2 of \cite{pruppacher2010microstructure} and also those produced by the E-AIM web-app \cite{wexler2002atmospheric}. For all regions the mid-residence-time in the region plus the time spent in all the previous regions is taken as the time duration for calculating the grown size of particles. The total time duration that the particles spend in the respiratory tract per each inhalation+exhalation maneuver is calculated as $60/f_R$, where $f_R$ is the respiration frequency per minute provided by the ICRP94 model. The time that particles spend in each region is then calculated by the distribution of the total respiration time according to the time constants provided by the ICRP94 deposition model for thoracic bronchial, bronchioles and alveolar regions. The particle residence times for the extrthoracic regions, which are not provided in the ICRP94 model, during inhalation or exhalation are assumed to be \SI{0.1}{\milli\second}.

The fractional ratio $f_d\left(d_0, \lambda_{I}(t), w(d_0, t), t\right)$ is one of the most challenging parameters in Eq. \ref{eqn:absorbed_dose_full}. This single parameter should include the fluid physics of how the exhale flow mixes with the environment and possible particles losses due to deposition on surfaces before reaching the susceptible. This parameter depends on the size of the exhaled particles, respiratory activity, shrinkage factor, advection distance/time from infectious to susceptible, room conditions (relative humidity, temperature, air flow, ventilation type), anatomical and physiological features of the infectious/susceptible individuals and whether or not the infectious individual is wearing a face mask (since it significantly affects the exhale flow). Due to the dynamical variability in room air conditions, it is extremely challenging (if not impossible) to fully determine $f_d$ even in a single scenario involving a single source. Even the most detailed simulations to date are carried out by \emph{assuming} the exhale flow behaves similar to a turbulent jet in a room with quiescent air \cite[e.g. see][and reference therein]{Chong_2021, Wang2021}. Therefore, in this study, we use a simplified theoretical formulation proposed for particle-laden jet flows \cite{Yang2020}, i.e. $f_d = a/(x\, tan(\alpha))$,  where $x$ is the distance between the source and the receptor, $a$ is the radius of the mouth (assuming a circular shape) and $\alpha$ is the exhale jet half-angle. For $x=$\SI{1}{\meter}, $a=$\SI{1.2}{\centi\meter} and $\alpha\sim$\SI{10}{\degree}, $f_d$ is approximately 5.6\%, which agrees well with the 4.9\% experimentally measured for \SI{0.77}{\micro\meter} particles by \cite{Liu2014}. For nose breathing, \citep{Gupta_2010} found average nose opening area of 0.56-0.71 \SI{}{\square\centi\meter} ($a=0.42-0.48$ \SI{}{\centi\meter}) and $\alpha\sim$\SI{11.5}{\degree}, where $f_d=2-3\% $ at a distance of \SI{1}{\meter}. For mouth breathing, \citep{Gupta_2010} found $a\sim$0.61-0.75 \SI{}{\centi\meter} and $\alpha\sim$\SI{17}{\degree}, where $f_d=2\%$ at a distance of \SI{1}{\meter}. For speaking, \citep{Gupta_2010} found average mouth opening of \SI{1.8}{\square\centi\meter}, which corresponds to $a\sim$\SI{0.76}{\centi\meter}. however, no information for $\alpha$ is presented. In order to be on the conservative side when calculating infection risk, we assume $a=$\SI{1.8}{\centi\meter} and $\alpha=$\SI{10}{\degree} to achieve $f_d=0.1$ at a distance of \SI{1}{\meter}. These values are used for all scenarios to calculate $f_d$.

For scenarios involving individuals with face masks, $f_{d}$ is expected to decay much faster with distance compared to those without face masks. 
Nonetheless, given the lack of a quantitative method to estimate the dilution factor and to not weaken our results by assuming an \emph{arbitrary value} for $f_{d}$, infection risk for masked individuals are calculated assuming $f_{d}=1$.
This, of course, leads to an upper bound with a large overestimation of the risk for masked individuals. Later we discuss possible implications of this assumption on the presented results.

\subsection*{Author contributions}
GB, BT, BH, OS, EB: experiment design, writing original draft and final text; BT, OS: conducting experiments with human subjects, GB, BT, OS, EB: intercomparison and analysis of particle spectrometer data, measurement of filter fabric penetration; GB and BH: analyses of the exhalation cloud vs. well-mixed room; GB, OS, EB: conducting particle decay experiments reported in Table 1; GB: calculation of infection risk.

\subsection*{Acknowledgments}
The authors would like to thank Udo Schminke and his team of the MPI-DS machine shop for providing us with mask holders, adaptors and other experiment-specific hardware. Special thanks to Prof. Simone Scheithauer (University Medicine G\"ottingen) for many helpful discussions in regards to SARS-CoV-2 and infectious aerosols. We also thank Dr. Mira P\"ohlker (MPI for Chemistry, Mainz) for allowing us to use her NS-SMPS after returning from the EUREC$^4$A research cruise. We thank her and Dr. Christopher P\"ohlker (also MPI for Chemistry) for the many helpful discussions on human aerosols.


\bibliography{Manuscript}

\end{document}